\documentclass{article}
\usepackage[T1]{fontenc}
\usepackage[utf8]{inputenc}
\usepackage{ismir}
\newcommand{\permission}{}
\usepackage{amsmath,amssymb,cite,url}
\usepackage{graphicx}
\usepackage{color}
\usepackage{booktabs}
\usepackage{multirow}
\usepackage{enumitem}

\makeatletter
\define@key{Gin}{alt}[]{}
\makeatother

\title{ARIMA: Reconstruction-Grounded Predictive Representation Learning for Symbolic Music}

\multauthor
{Mingyang Yao \hspace{1cm} Zhaoxiang Feng} {
University of California, San Diego \\
{\tt\small \{m5yao, zhf004\}@ucsd.edu}
}

\begin{document}

\maketitle

\begin{abstract}
Self-supervised learning for symbolic music has advanced largely through token-level pretraining, but such representations remain tied to tokenizer-specific sequences and often provide time-span-level embeddings only indirectly. In this paper, we propose ARIMA, a reconstruction-grounded latent predictive framework for symbolic music that learns compact window-based representations directly from data. ARIMA encodes each fixed-duration window into a continuous latent representation, trains a causal predictor with contrastive next-latent prediction, and grounds the encoder through structured reconstruction of music elements. This design preserves local musical details while modeling temporal progression across windows. We evaluate ARIMA on downstream tasks spanning various levels of music understanding. Results show that ARIMA is particularly efficient and effective on tasks involving harmonic, timing, and cross-performance retrieval, while remaining competitive with much larger baselines on other tasks. Ablations further show that next-latent prediction is essential for temporally integrated representations, and that structured reconstruction stabilizes latent learning without requiring explicit variance regularization. The code is here\footnote{https://github.com/AndyWeasley2004/symbolic\_music\_wm}
\end{abstract}

\section{Introduction}\label{sec:introduction}

Self-supervised learning (SSL) has become an important approach for learning transferable representations from unlabeled symbolic music. Recent works such as MidiBERT-Piano~\cite{chou2024midibertpiano}, MusicBERT~\cite{zeng2021musicbert}, PianoBART~\cite{liang2024pianobart}, and Aria~\cite{bradshaw2025aria} adapt masked language modeling or autoregressive objectives to MIDI tokenizations~\cite{le2025nlpmusic}, including REMI~\cite{huang2020remi}, CP~\cite{hsiao2021cpword}, OctupleMIDI~\cite{zeng2021musicbert} and other variants. In parallel, contrastive models such as CLaMP~\cite{wu2023clamp} learn joint representations between symbolic music and natural language, enabling semantic retrieval and zero-shot music classification. These studies demonstrate the value of large-scale SSL for music understanding and generation.

Despite this progress, most existing symbolic SSL remain closely tied to token-level or cross-modal alignment objectives. Token-based models learn from tokenizer-specific event sequences. While this is effective for generation and note-level classification, time-span-level representations are usually obtained only indirectly through pooling or task-specific adaptation. Moreover, the token counts corresponding to the same temporal span can vary substantially with note density, time resolution, and tokenizer design, making musical timings less explicit. Cross-modal models such as CLaMP provide higher-level semantic supervision, but their objective primarily captures language-associated musical semantics rather than the temporal evolution of music. This leaves a gap for self-supervised objectives that learn compact time-span-level representations while directly modeling progression across time.

A related direction has been explored through joint-embedding predictive architectures (JEPA)~\cite{assran2023ijepa,bardes2024vjepa,maes2026leworldmodel} under the world model domain. Instead of predicting the next frame as raw images or discrete tokens, JEPA-style models learn by predicting the next state latent from context and actions. Specifically, an encoder maps each input image to an embedding, while a predictor models causal dependencies among embeddings in latent space. This provides a natural analogy for symbolic music: local windows can be encoded as musical states, and self-supervised prediction across windows can encourage the model to learn how these states progress over time.

However, symbolic music also differs from vision and video in an important aspect. Low-level information such as pitch, onset patterns, sustain behavior, and velocity often carries direct analytical meaning in symbolic music. A purely non-reconstructive latent prediction objective may therefore learn high-level temporal abstractions while imposing limited pressure to preserve these musically important details in each window. This motivates a reconstruction-grounded predictive framework: the model should be able to compress, retaining enough details, and predict, keeping aware of what could come next.

Therefore, we propose ARIMA, an \textbf{A}utoregressive \textbf{R}epresentation learning framework for Symbol\textbf{i}c \textbf{M}usic \textbf{A}nalysis. It 1) encodes each fixed-duration window of music into a continuous latent using a window encoder supervised by pianoroll, chroma, and velocity reconstruction, and 2) models temporal progression with a causal predictor trained by contrastive next-latent prediction~\cite{oord2018cpc}, combining local elements with longer context modeling.

Our contributions are summarized as follows:
\begin{itemize}[leftmargin=*]
    \item We propose ARIMA, the first window-level symbolic music representation learning framework.
    \item ARIMA reaches the best results on approximately half of the evaluated tasks and remains competitive on the others, despite being substantially smaller than the comparable baselines.
    \item We analyze how ARIMA learns different features and the impact of variance regularization on latents.
\end{itemize}

\section{Related Work}\label{sec:related}

Our work intersects with SSL for symbolic music and joint-embedding predictive frameworks. We therefore review each area.

\subsection{Self-supervised learning for symbolic music.}
Masked language modeling (MLM) is the dominant pretraining paradigm for symbolic music. MidiBERT-Piano~\cite{chou2024midibertpiano} adapts BERT-style pretraining and explores the effects of CP~\cite{hsiao2021cpword} and REMI~\cite{huang2020remi} tokenizations on model performance. Later, MusicBERT~\cite{zeng2021musicbert} introduces OctupleMIDI tokenization for bar-level and multi-track modeling. Diverging from encoder-only models, PianoBART~\cite{liang2024pianobart} applies BART-style denoising autoencoding to symbolic music. Recently, M2BERT~\cite{wang2025m2bert} goes beyond standard NLP objectives and proposes music-specific pianoroll and chroma reconstructions from masked regions, which shows the potential in learning objectives following the inductive bias of music. In addition to the MLM family, Aria~\cite{bradshaw2025aria} scales autoregressive pretraining and contrastive learning to a 631M-parameter model trained on $\sim$60{,}000\,hours of transcribed piano MIDI. These approaches all operate over tokens, whereas our framework operates at the window level.

\subsection{Joint-embedding predictive architectures.}
Latent-space sequence modeling has been substantially explored through JEPA, which learns by predicting representations rather than raw observations. I-JEPA~\cite{assran2023ijepa} predicts masked image-block embeddings from visible context, and V-JEPA~\cite{bardes2024vjepa} extends this formulation to spatio-temporal prediction in video. More recently, LeWorldModel~\cite{maes2026leworldmodel} frames JEPA as an end-to-end latent world model, using next-embedding prediction to learn the causal effect of actions applied on the current state embedding. Relatedly, Contrastive Predictive Coding (CPC)~\cite{oord2018cpc} provides a contrastive learning paradigm for supervising future latent states with an InfoNCE objective instead of objectives that direct matches the prediction to the targets.

We follow this latent predictive design but adapt it to symbolic piano performance. Unlike general JEPA-style world models, ARIMA additionally grounds the encoder with structured symbolic reconstruction, placing it between contrastive predictive learning, JEPA-style latent prediction, and analysis-oriented symbolic music representation learning.

\section{Method}\label{sec:method}

\begin{figure*}[t]
\centering
\includegraphics[width=\linewidth]{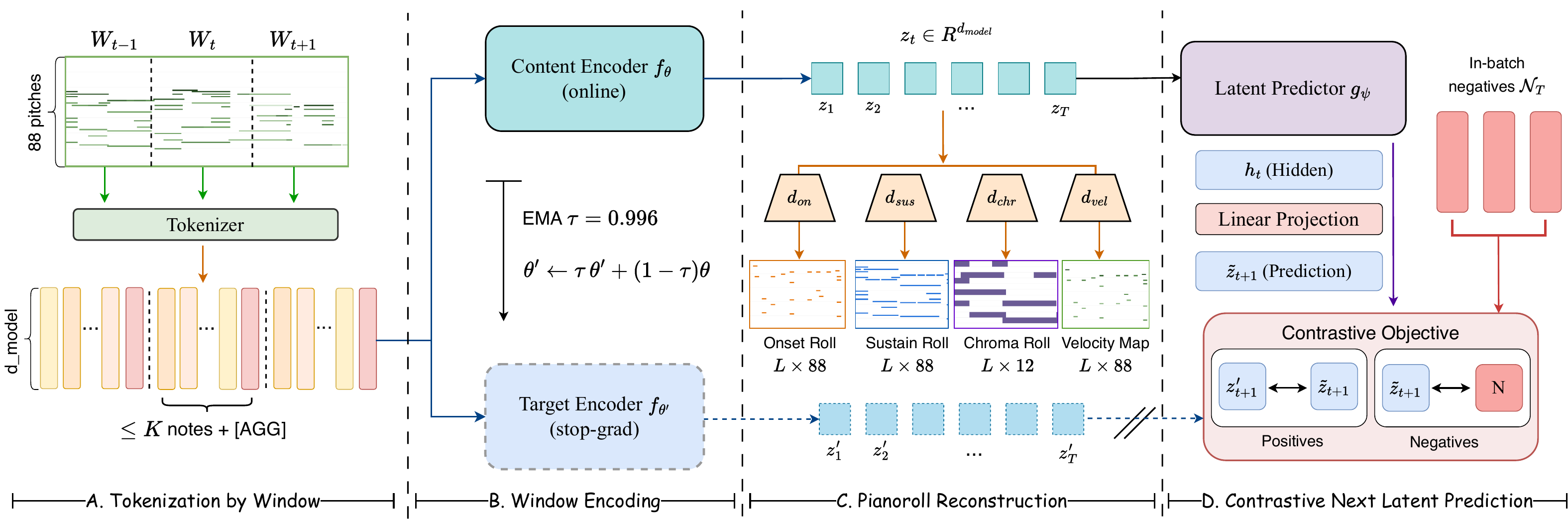}
\vspace{-0.5cm}
\caption{The architecture and training dynamics of ARIMA, from the tokenization, window encoding (left), to structured pianoroll reconstruction and contrastive next latent prediction (right)}
\vspace{-0.3cm}
\label{fig:architecture}
\end{figure*}

Figure~\ref{fig:architecture} illustrates the preprocessing pipeline, model backbone, and main objectives in different modules. We introduce the input representation, model components, and training objectives in the specification below.

\subsection{Input Representation}\label{subsec:input}

\subsubsection{Window partitioning and note encoding.}
Given a piece of symbolic music, we partition it into a sequence of $T$ windows $\{W_1, \ldots, W_T\}$ of length $\Delta$ seconds, where each window $W_t$ contains at most $K$ notes. Then, each note is encoded as a tuple
\begin{equation}
n = \bigl(p,\; v,\; o,\; \delta,\; \iota,\; \phi_{\text{prev}},\; \phi_{\text{next}}\bigr),
\end{equation}
where $p \in \{0, \ldots, 127\}$ is MIDI pitch, $v \in \{0, \ldots, 127\}$ is velocity, $o$ and $\delta$ are onset time and duration in milliseconds relatjive to the window start, $\iota$ is the inter-onset interval (IOI), and $\phi_{\text{prev}}, \phi_{\text{next}} \in \{0, 1\}$ indicate whether the note is tied from the previous window or into the next. Notes crossing a boundary appear in both adjacent windows with truncated durations. Specifically, $\phi_{\text{next}}=1$ in the first window, and $\phi_{\text{prev}}=1$ with $o=0$ in the second, so no musical event is lost at the partition.

\subsubsection{Note embeddings.}
Each note $n$ is mapped to an embedding by summing learned categorical embeddings for the discrete fields $(p, v, \phi_{\text{prev}}, \phi_{\text{next}})$ and sinusoidal encodings~\cite{vaswani2017attention} of the continuous timing fields $(o, \delta, \iota)$. For window $t$, we denote the resulting sequence of note embeddings as $\mathcal{W}_t$. Although the input uses the exact categorical velocity value, the velocity reconstruction target is rescaled to $[0,1]$, optimizing as an ordered intensity value.

\subsection{Model}\label{subsec:arch}

ARIMA learns two window-level representations: a \emph{content latent} $z_t \in \mathbb{R}^d$ produced by an encoder, and a \emph{temporal hidden state} $h_t \in \mathbb{R}^d$ produced by a causal predictor.

\subsubsection{Window encoder.}
The encoder $f_\theta : \mathcal{W} \to \mathbb{R}^d$ is a encoder that operates independently on the embedded notes in each window with a learnable aggregation token. Its final aggregation state is the content latent:
\begin{equation}
z_t = f_\theta(\mathcal{W}_t).
\end{equation}

\subsubsection{Causal predictor.}
The predictor $g_\psi : \mathbb{R}^{T \times d} \to \mathbb{R}^{T \times d}$ is a causal decoder that maps the sequence of content latents to temporal hidden states,
\begin{equation}
(h_1, \ldots, h_T) = g_\psi(z_1, \ldots, z_T),
\end{equation}
where each $h_t$ depends only on $z_{1:t}$. A linear head $\pi_\phi$ then produces the next-latent prediction $\hat{z}_{t+1} = \pi_\phi(h_t) \in \mathbb{R}^d$.

\subsubsection{EMA target encoder.}
Following~\cite{grill2020byol}, a stop-gradient copy $f_{\theta'}$ of the online window encoder is maintained by an exponential moving average (EMA),
\begin{equation}
\theta' \leftarrow \tau\, \theta' + (1 - \tau)\, \theta,
\end{equation}
with momentum $\tau \in [0, 1)$. This target encoder produces the stable contrastive target $z'_t = f_{\theta'}(W_t)$.

\subsubsection{Reconstruction heads.}
Four lightweight MLPs map the \emph{content latent} $z_t$ to structured musical targets: $d_{\text{on}}, d_{\text{sus}}: \mathbb{R}^d \to [0,1]^{L \times 88}$ produce the onset $O$ and sustain $S$ rolls, $d_{\text{chr}}: \mathbb{R}^d \to [0,1]^{L \times 12}$ produces the chromaroll $C$, and $d_{\text{vel}}: \mathbb{R}^d \to \mathbb{R}^{L \times 88}$ produces the velocity map $V$, where $L$ is the number of temporal bins per window.

\subsection{Training Objectives}\label{subsec:objectives}

The training loss combines contrastive next-latent prediction, binary reconstruction, and velocity regression:
\begin{equation}
\mathcal{L} = \lambda_{\text{pred}} \mathcal{L}_{\text{pred}} + \lambda_{\text{recon}} \mathcal{L}_{\text{recon}} + \lambda_{\text{vel}} \mathcal{L}_{\text{vel}}.
\end{equation}

\subsubsection{Next-latent prediction.}
Let $\tilde{z}_{t+1} = \hat{z}_{t+1} / \|\hat{z}_{t+1}\|_2$ and $\tilde{z}'_{t+1} = z'_{t+1} / \|z'_{t+1}\|_2$ denote the $\ell_2$-normalized prediction and target. The predictor is trained with an InfoNCE loss~\cite{oord2018cpc} at each position $t$,
\begin{equation}\label{eq:infonce}
\mathcal{L}_{\text{pred}}^{(t)} = -\log \frac{\exp\,\!\bigl(\tilde{z}_{t+1}^{\top} \tilde{z}'_{t+1} / \tau\bigr)}{\sum_{\bar{z} \in \mathcal{N}_t \cup \{\tilde{z}'_{t+1}\}} \exp\,\!\bigl(\tilde{z}_{t+1}^{\top} \bar{z} / \tau\bigr)},
\end{equation}
with temperature $\tau > 0$. The negative set $\mathcal{N}_t$ consists of all other normalized target latents in the batch. Pitch-transposed copies of the \emph{same} window are excluded from $\mathcal{N}_t$, while same-piece, different-window views are treated as hard negatives and doubled their weights. This encourages transposition invariance in $z$ while avoiding piece-level fingerprinting.

\subsubsection{Pianoroll and chroma reconstruction.}
Given ground-truth onset pianoroll $O_t$, sustain pianoroll $S_t$, and chromaroll $C_t$, the reconstruction loss aggregates three terms,
\begin{equation}
\mathcal{L}_{\text{recon}} = \text{OWBCE}(\hat{O}, O) + \text{BCE}_\rho(\hat{S}, S) + \text{BCE}_\rho(\hat{C}, C),
\end{equation}                                             where $\text{BCE}_\rho$ is binary cross-entropy with positive-class weight $\rho \ge 1$. The onset-weighted variant $\text{OWBCE}$ adapts the soft-onset idea common in piano transcription~\cite{kong2021piano}, which smooths continuous onset times over neighboring bins to absorb the quantization jitter from mapping real-valued onsets onto a fixed grid; unlike prior work that softens the \emph{target}, we keep $O$ binary and upweight the \emph{loss} near each onset. Concretely, we convolve $O$ along the time axis with a discrete sine window $k$ of full width $2w{+}1$ (peak $1$ at the center, zero at both edges) to obtain $\tilde{O} = \min(k * O, 1) \in [0, 1]^{L \times 88}$, and weight each cell by
\begin{equation}
W[l, i] = 1 + \alpha\, \tilde{O}[l, i],
\end{equation}
so a bin at an onset contributes up to $(1+\alpha){\times}$ the loss of a bin far from any onset, which is a soft temporal margin without altering the target.  

\subsubsection{Velocity regression.}
We use standard MSE to supervise velocity reconstruction:
\begin{equation}
\mathcal{L}_{\text{vel}} = \frac{1}{|\Omega_t|} \sum_{(l, k) \in \Omega_t} \bigl(\hat{V}_t[l, k] - V_t[l, k]\bigr)^2,
\end{equation}
and the loss is restricted to onset positions $\Omega_t = \{(l, k) : O_t[l, k] = 1\}$.

\subsubsection{Augmentation.}
Each window receives pitch transposition within $\pm 3$ semitones and velocity jitter within $\pm 5$ units. Transposed views participate in $\mathcal{N}_t$ as hard negatives, except for augmented copies of the same window as described above.

\section{Experiments}\label{sec:experiments}

\subsection{Data and Model Setting}\label{subsec:training}

The window encoder is a 4-layer Transformer encoder, and the predictor is a 6-layer causal Transformer decoder with learned relative position bias. Both use $d = 512$, 8 heads, and an FFN dimension of 2048. Training is regularized with dropout and stochastic depth~\cite{huang2016stochasticdepth} probability of 0.1. Each window spans $\Delta = 2$s with at most $K = 100$ notes and $L = 32$ temporal bins for reconstruction, and the predictor uses $T = 60$ windows of context, corresponding to 120s effective duration. The EMA momentum of the target encoder is $\tau = 0.996$. In total, ARIMA has $\sim$38M trainable parameters. The training data consists of $\sim$15{,}000 piano performances from ATEPP~\cite{zhang2022atepp}, POP1K7~\cite{hsiao2021cpword}, POP909~\cite{wang2020pop909}, and EMOPIA~\cite{hung2021emopia}.

\subsection{Training Configuration}

The model is trained end-to-end using AdamW~\cite{loshchilov2019adamw} with $\beta = (0.9, 0.95)$ and weight decay $0.05$. The learning rate is linearly warmed up for $2{,}000$ steps to a peak of $2 \times 10^{-4}$, then linearly decayed to 25\% of the peak over 42{,}000 steps. We use mixed-precision BF16 training and set the maximum gradient norm to $1.0$.

For reconstruction, we set the OWBCE kernel half-width to 3 with $\alpha = 12$, the $\text{BCE}_\rho$ positive weight to $\rho = 3.0$, $\lambda_{\text{recon}} = 1.0$, and $\lambda_{\text{vel}} = 5.0$. For contrastive latent prediction, we use InfoNCE temperature $\tau_I = 0.07$ and warm up the contrastive term weight from $0$ to $\lambda_{\text{pred}} = 0.25$ over the first 8{,}000 steps, allowing the encoder to establish a stable reconstruction-anchored latent space before the predictor begins shaping it.

\subsection{Downstream Tasks}\label{subsec:tasks}

We evaluate on 9 tasks spanning six datasets. The first 6 are basic understanding tasks: \emph{composer classification} under two label sets (top 21- and 16-composer on ATEPP~\cite{zhang2022atepp} and the \emph{8 pianists} of Pianist8~\cite{chou2021pianist8}), \emph{performer identification} with top 32- and 16-performer on ATEPP, four-quadrant \emph{emotion classification} on EMOPIA~\cite{hung2021emopia}, \emph{key estimation} on human-annotated POP909~\cite{wang2020pop909, yao2026bachi}, and ordinal Henle-scale difficulty estimation on CIPI~\cite{ramoneda2024cipi}. 

The remaining 3 are ASAP-based tasks~\cite{foscarin2020asap} targeting performance structure: \emph{piece-level Inter-Onset Interval (IOI) regression}, \emph{performer verification}, and \emph{cross-performance retrieval}. The IOI task predicts the per-piece mean of $\log(\mathrm{IOI}_\text{perf}/\mathrm{IOI}_\text{score})$ across annotated beats, i.e., the log tempo ratio relative to the reference score. Performer verification asks whether two performances by the same performer are closer in cosine space than two performances by different performers, with positive (same-performer / different-score) and negative (different-performer) pairs sampled at a 1:5 ratio. Finally, cross-performance retrieval queries whether each performance's nearest cosine neighbor is a different rendition of the same score, reported as R@1. 

For all classification tasks, we perform mean pooling on models producing representation sequences, such as MLM-based models and ARIMA, and use the produced embedding directly for models providing global representation, including public Aria embedding and our replica. Using these embeddings, we train a two-layer MLP with 256 hidden units under 5-fold cross-validation, and report the macro F1 score. On CIPI, we applied an ordinal regression probe with its pre-defined data splits and reported as Spearman correlation~$\rho$, which measures if the model orders pieces by difficulty in the same order as the ground-truth ranking. For probing tasks mentioned above, we evaluate on both the content latent $z$, and the predictor latent $h$, and verification and retrieval are computed directly from cosine similarities without a learned probe.

\subsection{Baselines and Ablations}\label{subsec:baselines}

\subsubsection{Baselines.}
We compare against four pretrained models and one replicated model. For pretrained models with open-source weights, we use \textbf{MidiBERT-Piano}~\cite{chou2024midibertpiano}, \textbf{PianoBART}~\cite{liang2024pianobart}, \textbf{M2BERT}~\cite{wang2025m2bert}, and Aria variants (\textbf{Aria-base} and \textbf{Aria-embedding})~\cite{bradshaw2025aria}. To isolate paradigm from scale, we additionally train one Aria replica at the $\sim$100M size with the same corpus used by ARIMA and the same upstream architecture, optimization, and augmentation policy. To reduce the size, the model is set to $8$ layers with $d_\text{model}=768$.

We interpret comparisons against the public Aria models as foundation-scale references rather than controlled comparisons, since they differ substantially from all other models in parameter count and training data size. The 100M replica provides a more controlled comparison of the paradigm under a closer model scale. All baselines take performance MIDI as input and produce a mean-pooled representation across the full piece.

\subsubsection{Ablation variants.}
We conduct an ablation study on four variants: (1) adding VICReg~\cite{bardes2022vicreg}, which is effective for preventing representation collapse~\cite{sobal2025learning}; (2) removing latent prediction and using the encoder only; (3) using a combined pianoroll as the reconstruction target instead of disentangled onset and sustain rolls; and (4) the full ARIMA model.

\section{Results}\label{sec:results}

\begin{table*}[t]
\centering
\footnotesize
\setlength{\tabcolsep}{2.5pt}
\begin{tabular*}{\linewidth}{@{\extracolsep{\fill}} l r cc c cc ccc c cc @{}}
\toprule
& & \multicolumn{8}{c}{\textbf{Content Probing}} & \textbf{Temporal} & \multicolumn{2}{c}{\textbf{Similarity}} \\
\cmidrule(lr){3-10} \cmidrule(lr){11-11} \cmidrule(lr){12-13}
& & \multicolumn{2}{c}{\textbf{Composer}} & & \multicolumn{2}{c}{\textbf{Performer}} & & & & & & \\
\cmidrule(lr){3-4} \cmidrule(lr){6-7}
\textbf{Model} & \textbf{\# Params} & \textbf{21-cls} & \textbf{16-cls} & \textbf{Pianist8} & \textbf{32-cls} & \textbf{16-cls} & \textbf{Emotion} & \textbf{Key} & \textbf{Difficulty} & \textbf{IOI} & \textbf{Verification} & \textbf{Retrieval} \\
 & & ($F_1$) & ($F_1$) & ($F_1$) & ($F_1$) & ($F_1$) & ($F_1$) & ($F_1$) & ($\rho$) & ($R^2$) & (AUC) & ($R@1$) \\
\midrule
MidiBERT-Piano     & 110M & 82.1 & 86.5 & 68.0 & 25.4 & 36.6 & 68.8 & 77.5 & .674 & .674 & .466 & 93.8 \\
PianoBART Encoder  & 102M & 84.8 & 89.0 & 77.6 & 46.0 & \underline{59.6} & 71.7 & 76.7 & \textbf{.772} & .605 & .676 & 92.8 \\
M2BERT            & 61M  & \underline{90.4} & \underline{92.3} & 74.0 & 39.7 & 51.7 & 70.8 & \underline{81.7} & .722 & \underline{.795} & .641 & \underline{98.7} \\
\midrule
Aria-base-100M       & 103M & 89.5 & 92.0 & \textbf{84.1} & \textbf{55.2} & \textbf{71.9} & \textbf{74.5} & 77.7 & .697 & .749 & .695 & 98.3 \\
Aria-embedding-100M & 89M  & 77.7 & 83.7 & 71.0 & 30.1 & 40.2 & 68.0 & 46.8 & .660 & .593 & .698 & 90.8 \\
\midrule
ARIMA ($z$, ours)  & 38M  & 86.5 & 90.6 & 75.7 & 38.4 & 50.7 & 68.8 & \textbf{82.7} & .726 & .738 & \underline{.768} & \textbf{99.0} \\
ARIMA ($h$, ours)  & 38M  & \textbf{92.1} & \textbf{94.9} & \underline{79.0} & \underline{47.2} & 59.3 & \underline{74.0} & 81.0 & \underline{.734} & \textbf{.804} & \textbf{.774} & \underline{98.6} \\
\midrule
Aria-base (reference)         & 659M & 43.8 & 49.6 & 44.2 & 17.9 & 26.2 & 48.9 & 36.9 & .591 & .173 & .442 & 73.2 \\
Aria-embedding (reference)    & 632M & 95.2 & 96.0 & 90.8 & 65.5 & 78.0 & 80.6 & 9.9  & .608 & .672 & .905 & 86.7 \\
\bottomrule
\end{tabular*}
\caption{Downstream evaluation results. \textbf{Bold} / \underline{underline}: best / second-best among comparable-scale models ($\leq 110$M); the public Aria models are foundation-scale references, excluded from the ranking.}
\vspace{-3mm}
\label{tab:main}
\end{table*}

\subsection{Main Results}\label{subsec:main_results}

\subsubsection{Comparable-scale baselines.}
Table~\ref{tab:main} summarizes performance across all tasks. Among comparable-scale models, ARIMA is consistently competitive despite having the smallest parameter count. The predictor representation $h$ obtains the best results on composer classification, inter-onset interval regression, and performer verification, and ranks second on Pianist8, 32-class performer identification, difficulty estimation, and cross-performance retrieval. The encoder representation $z$ is strongest on key estimation and cross-performance retrieval, suggesting that the reconstruction-grounded window latent preserves pitch-class and content-similarity information particularly well. The $z$ and $h$ from our ARIMA collectively achieved state-of-the-art on 5 out of 9. The main weakness of ARIMA lies in dynamic understanding, which makes its pianist 8, performer and emotion classification score the second compared with the token-level autoregressive model. Notably, the two strongest models on key estimation, ARIMA-$z$ and M2BERT, both use chroma reconstruction during pretraining, which provides direct supervision on harmony content. Similarly, IOI regression benefits from explicit IOI features in the input encoding, though it's not supervised in reconstruction. This also flags the crucial role of inductive-bias alignment.

\subsubsection{Complementarity of $z$ and $h$.}
The two ARIMA representations exhibit complementary behavior. Since $z_t$ is computed from a single window, it is well-suited to tasks that depend on local content, such as key estimation and cross-performance retrieval. In contrast, $h_t$ aggregates causal history through the predictor, making it more effective for global tasks, including style-, performer-, emotion-, and IOI -related probes that benefit from longer-range context. This separation supports the intended design: $z$ captures local musical content, while $h$ captures temporally integrated structure.

\subsubsection{Foundation-scale references.}
The public Aria-embedding model remains the strongest on most tasks, which is expected given its substantially larger scale and training corpus. However, its performance is not uniformly dominant: it is much weaker on key and difficulty estimation, IOI regression, and cross-performance retrieval. This pattern aligns with the design of its contrastive finetuning, where two independently augmented random slices from the same MIDI file are treated as positives. Because the augmentations include transposition, tempo scaling, and velocity perturbation, the objective encourages invariance to factors needed for absolute pitch-center estimation, local timing-ratio prediction, and score-level identity across performances~\cite{bradshaw2025aria}. The retrieval gap is especially interpretable because Aria's positive pairs come from the same file, whereas retrieval tasks require different performances of the same score to be close.

In contrast, ARIMA preserves local chroma, pianoroll, velocity, and temporal progression through reconstruction and next-latent prediction, which better matches the mixture of local and global analysis tasks. Within the controlled 100M-scale replicas, the base variant is also markedly stronger than the embedding variant across most probes, further suggesting that contrastive piece-level embedding geometry can discard details useful for fine-grained symbolic analysis, and this issue will become serious without abundant training data.

Overall, the results indicate that a compact window-level model can match or exceed larger token-level baselines on several symbolic MIR tasks, especially those involving harmonic content, temporal behavior, and similarity-based retrieval.

\subsection{Ablation Study}\label{subsec:ablation}

\begin{table*}[t]
\centering
\footnotesize
\setlength{\tabcolsep}{2.5pt}
\begin{tabular*}{\linewidth}{@{\extracolsep{\fill}} l l cc c cc ccc c cc @{}}
\toprule
& & \multicolumn{8}{c}{\textbf{Content Probing}} & \textbf{Temporal} & \multicolumn{2}{c}{\textbf{Similarity}} \\
\cmidrule(lr){3-10} \cmidrule(lr){11-11} \cmidrule(lr){12-13}
& & \multicolumn{2}{c}{\textbf{Composer}} & & \multicolumn{2}{c}{\textbf{Performer}} & & & & & & \\
\cmidrule(lr){3-4} \cmidrule(lr){6-7}
\textbf{Variant} & \textbf{Rep.} & \textbf{21-cls} & \textbf{16-cls} & \textbf{Pianist8} & \textbf{32-cls} & \textbf{16-cls} & \textbf{Emotion} & \textbf{Key} & \textbf{Difficulty} & \textbf{IOI} & \textbf{Verification} & \textbf{Retrieval} \\
 & & ($F_1$) & ($F_1$) & ($F_1$) & ($F_1$) & ($F_1$) & ($F_1$) & ($F_1$) & ($\rho$) & ($R^2$) & (AUC) & ($R@1$) \\
\midrule
\multirow{2}{*}{ARIMA (final)}
  & $z$  & 86.5 & 90.6 & 75.7 & 38.4 & 50.7 & 68.8 & 82.7 & .726 & .738 & .768 & 99.0 \\
  & $h$  & \underline{92.1} & \underline{94.9} & \textbf{79.0} & \textbf{47.2} & \textbf{59.3} & \textbf{74.0} & 81.0 & .734 & \underline{.804} & .774 & 98.6 \\[3pt]
\multirow{2}{*}{$+$VICReg}
  & $z$ & 87.0 & 89.5 & 74.5 & 35.9 & 47.0 & 68.1 & 82.0 & .736 & .731 & .798 & 99.1 \\
  & $h$ & 91.7 & \textbf{95.2} & \underline{78.6} & \underline{46.2} & \textbf{59.3} & 72.0 & 81.9 & .739 & .797 & .772 & 98.7 \\[3pt]
\multirow{1}{*}{$-$Prediction}
  & $z$ & 81.4 & 86.0 & 69.0 & 27.8 & 39.7 & 66.0 & \underline{83.4} & .719 & .639 & .792 & 98.7 \\
  & $h$ & --- & --- & --- & --- & --- & --- & --- & --- & --- & --- & --- \\[3pt]
\multirow{2}{*}{$-$Onset/Sustain}
  & $z$ & 87.0 & 90.5 & 74.6 & 35.4 & 48.2 & 68.7 & \textbf{83.7} & \underline{.740} & .741 & \underline{.807} & \textbf{99.5} \\
  & $h$ & \textbf{92.4} & \textbf{95.2} & 78.4 & 44.8 & \underline{57.0} & \underline{73.0} & 81.9 & \textbf{.745} & \textbf{.818} & \textbf{.810} & \textbf{99.7} \\
\bottomrule
\end{tabular*}
\caption{Ablation study on ARIMA variants. Performance of all tasks is reported. The no-prediction variant has no $h$ result because the predictor is inactive. \textbf{Bold} / \underline{underline}: best / second-best per column across all rows.}
\vspace{-3mm}
\label{tab:ablation}
\end{table*}

\subsubsection{Prediction and representation quality.}
Table~\ref{tab:ablation} reports the ablation results. Contrastive next-latent prediction is the most important ablated component. Removing it not only eliminates the predictor latent $h$, but also substantially weakens the content latent $z$ across classification and temporal tasks. Although there is a marginal improvement on content-focused tasks like key and retrieval, other style probing tasks downgrade substantially. This indicates that the predictive objective is critical for helping the window encoder organize local windows into temporally integrated representations.

\subsubsection{Variance regularization.}
Adding VICReg~\cite{bardes2022vicreg} produces only marginal downstream changes. It slightly improves a few metrics but does not consistently improve either $z$ or $h$. Since the non-regularized model does not collapse and remains competitive, we omit VICReg from the final configuration.

\subsubsection{Onset--sustain disentanglement.}
Merging onset and sustain reconstruction leads to mixed results. The merged variant improves several similarity and temporal metrics, whereas the disentangled final model is stronger on Pianist8, performer identification, and emotion classification. We retain separate onset and sustain heads because they provide a more explicit attack-versus-hold decomposition and improve internal temporal interpretability. Thus, onset--sustain disentanglement is best interpreted as a structurally motivated and inductive-bias-aligned design rather than a uniformly superior choice on every downstream metric.

\subsection{Discussion}\label{subsec:discussion}

\subsubsection{Latent Geometry and VICReg}\label{subsec:vicreg_geometry}

\begin{figure}[t]
\centering
\includegraphics[width=\columnwidth]{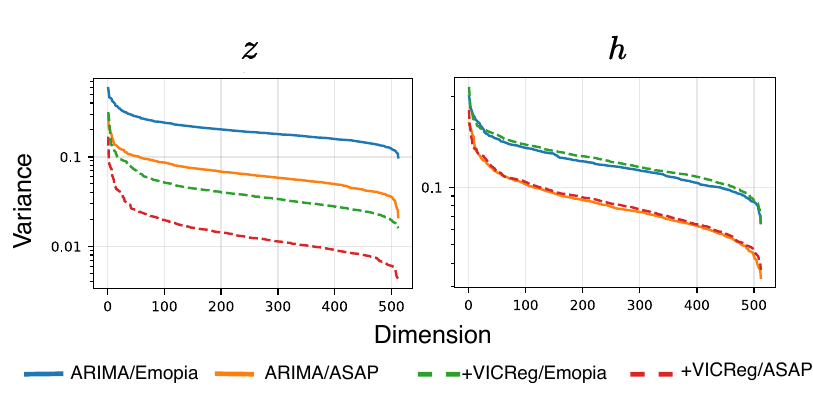}
\caption{Descending sorted per-dimension variance of content latent $z$ and predictor latent $h$ after piece-level mean pooling on EMOPIA and ASAP.}
\label{fig:var_per_dim}
\end{figure}

Although VICReg has little effect on downstream accuracy, it changes the geometry of the encoder latent space. As shown in Figure~\ref{fig:var_per_dim}, VICReg substantially increases the per-dimension variance of $z$, making the encoder representation more isotropic. However, this geometric change does not translate into consistently improved downstream performance. Meanwhile, the predictor representation $h$ is even less affected, which suggests the causal predictor and the normalization before the contrastive objective largely determine the geometry of $h$, while VICReg mainly reshapes the marginal statistics of $z$. These observations indicate that structured reconstruction of music elements already provides sufficient anti-collapse pressure, and explicit variance regularization is not necessary.


\subsubsection{Relation to JEPA-style latent prediction.}
A natural alternative is to train ARIMA as a more direct JEPA-style model, where the predictor regresses future latents without reconstruction. In preliminary experiments, however, variants using normalized $\ell_1$ latent regression or non-reconstructive next-latent prediction did not yield stable checkpoints, either diverging early or collapsing rapidly to near-trivial prediction losses. We do not report these as formal ablations, since they were not stable enough for evaluation. Nevertheless, they motivate a key design distinction between ARIMA and pure JEPA-style objectives. First, in action planning via vision, the current state and action produce a deterministic next state, but ``next-window" of music can have multiple choices, which is likely the trigger causing the direct regression on target to fail. Secondly, dense pixel-level redundancy and masking strategies on visions can provide strong learning signals even without reconstructing at the pixel level, but in symbolic music, the input is sparse, event-based, with important low-level information such as pitch, onset, durations, and velocity. Therefore, structured reconstruction on music elements acts as a grounding mechanism, preventing the latent space from becoming arbitrary while still allowing the predictor to learn progression abstraction.

\section{Conclusion}\label{sec:conclusion}
In this paper, we propose ARIMA, a reconstruction-grounded latent predictive framework for symbolic music representation learning. ARIMA encodes fixed-duration music windows into compact continuous latents, grounds these latents through structured reconstruction, and models temporal progression with contrastive next-latent prediction. Experiments across nine downstream tasks show that ARIMA learns effective representations for harmonic, timing, and cross-performance analysis while remaining substantially smaller than several comparable baselines. Ablations show that next-latent prediction is important for organizing window representations across time, while structured reconstruction provides sufficient anti-collapse pressure without explicit variance regularization. Future work includes scaling ARIMA to larger and more diverse symbolic corpora, extending the framework to a structure-aware hierarchical SSL framework, and apply to music generation and editing. All codes and model weights will be released after upon acceptance.

\section{Acknowledgement}
Generative AI tools assisted with manuscript revision and proofreading. All scientific decisions, experimental designs, figure drawing, results analysis and discussion were completely made by humans. 

\bibliography{ISMIRtemplate}

\end{document}